Matter in Space

R J Potton

Magdalene College, Cambridge

**Abstract**

A new linear mapping of the linear vector space (LVS) of the octonions is suggested as an approach to the co-ordinatization of space-time. This approach resolves some perplexing issues concerning the validity of certain pre-metric notions of orthogonality and mirror symmetry. It makes explicit the nature of the six-dimensional symplectic space that was alluded to earlier by the author and shows that the null space component of the extended space can be related to ideas that were set out many years ago by H J S Smith.

rjp73@cam.ac.uk

## 1 Introduction

Following Hamilton's introduction of quaternions and the subsequent ubiquity of vector products it was natural to consider orthogonal Cartesian frames for the location of matter in space [1,2]. Recently it weighs in on us that properties of the higher Cayley Dickson algebra of octonions has properties that have physical relevance. The suggestion here is that the combinatorial properties of scalars, vectors, bivectors and trivectors are essentially as given by the three types of multiplication described in the Appendix. This leaves open, however, an issue of dimensional reduction whereby the unique properties of the octonions should constrain the vector space structure that is appropriate for the understanding of how objects in the world behave. In the light of past theories and with remarks in [3] in mind, it seems natural to examine the eight-dimensional linear vectors space (LVS) of the octonions from the point of view that it can be seen as the direct sum of a six-dimensional symplectic space and a two dimensional representation of space-time.

Smith [4] has shown that symmetry arguments can be applied to lines and planes aligned in different directions in three-dimensional space. It turns out that the interconnections of his pre-metric [5] *symmetries of aspect* are related to null spaces in octonion space, the existence of which imply that there is a mapping of the octonion LVS to an image space of lower dimension depending on lattice symmetry. This image space is spanned by hypercomplex numbers and also connect with the octonion trivector unit $\mathbf{e}_7$.

## 2 Symplectic space

The multiplicative and associative properties of octonion units are given by lines in the Fano diagram which is a projective plane. As such, the possibility of a metric within the octonion linear vector space (LVS) remains an open question [6].

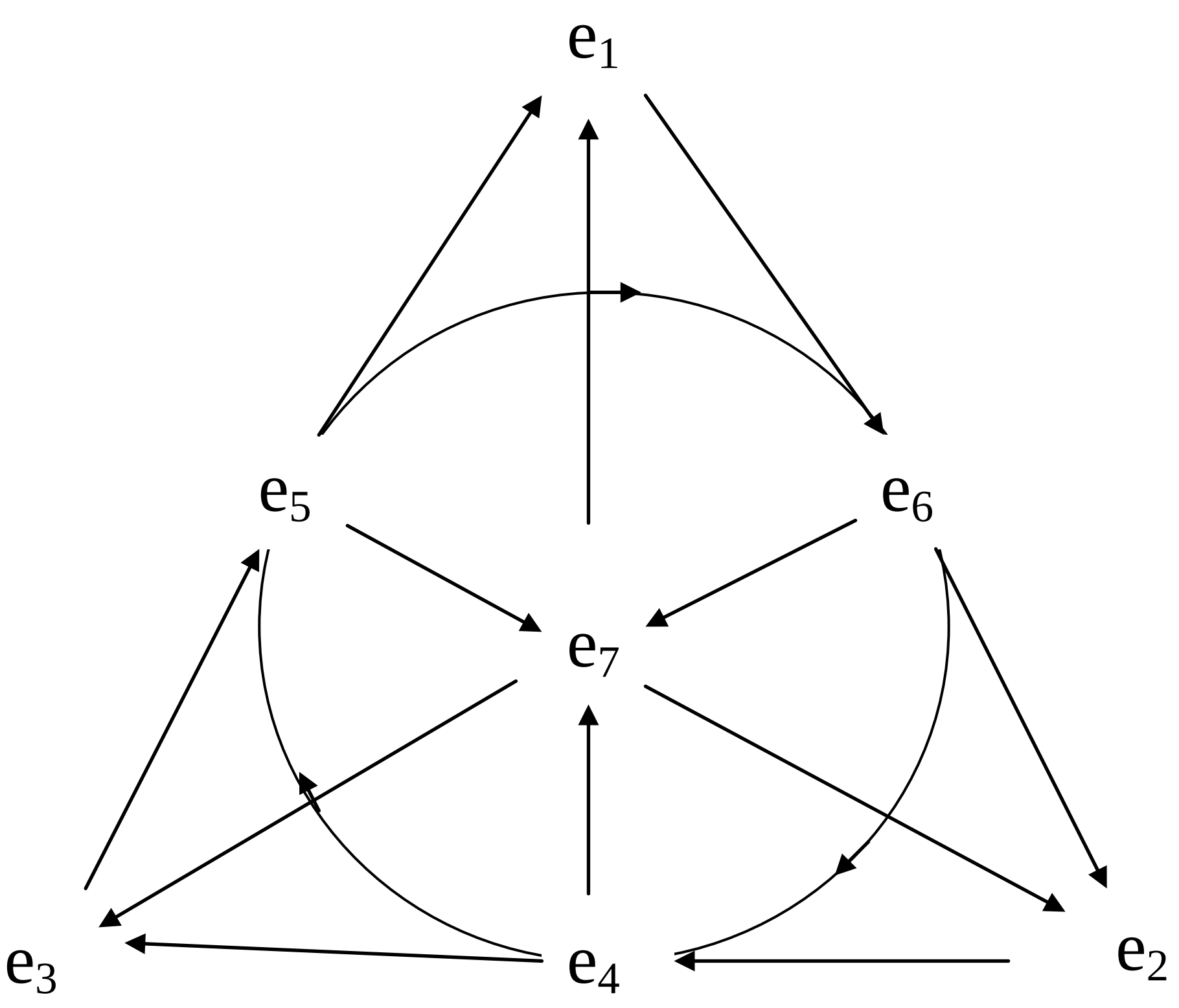

Figure 1 The Fano plane of a particular octonion multiplication table

| | $\mathbf{e}_0$ | $\mathbf{e}_1$ | $\mathbf{e}_2$ | $\mathbf{e}_3$ | $\mathbf{e}_4$ | $\mathbf{e}_5$ | $\mathbf{e}_6$ | $\mathbf{e}_7$ |
|---|---|---|---|---|---|---|---|---|
| $\mathbf{e}_0$ | **1** | $\mathbf{e}_1$ | $\mathbf{e}_2$ | $\mathbf{e}_3$ | $\mathbf{e}_4$ | $\mathbf{e}_5$ | $\mathbf{e}_6$ | $\mathbf{e}_7$ |
| $\mathbf{e}_1$ | $\mathbf{e}_1$ | **-1** | $-\mathbf{e}_6$ | $\mathbf{e}_5$ | $\mathbf{e}_7$ | $-\mathbf{e}_3$ | $\mathbf{e}_2$ | $-\mathbf{e}_4$ |
| $\mathbf{e}_2$ | $\mathbf{e}_2$ | $\mathbf{e}_6$ | **-1** | $-\mathbf{e}_4$ | $\mathbf{e}_3$ | $\mathbf{e}_7$ | $-\mathbf{e}_1$ | $-\mathbf{e}_5$ |
| $\mathbf{e}_3$ | $\mathbf{e}_3$ | $-\mathbf{e}_5$ | $\mathbf{e}_4$ | **-1** | $-\mathbf{e}_2$ | $\mathbf{e}_1$ | $\mathbf{e}_7$ | $-\mathbf{e}_6$ |
| $\mathbf{e}_4$ | $\mathbf{e}_4$ | $-\mathbf{e}_7$ | $-\mathbf{e}_3$ | $\mathbf{e}_2$ | **-1** | $\mathbf{e}_6$ | $-\mathbf{e}_5$ | $\mathbf{e}_1$ |
| $\mathbf{e}_5$ | $\mathbf{e}_5$ | $\mathbf{e}_3$ | $-\mathbf{e}_7$ | $-\mathbf{e}_1$ | $-\mathbf{e}_6$ | **-1** | $\mathbf{e}_4$ | $\mathbf{e}_2$ |
| $\mathbf{e}_6$ | $\mathbf{e}_6$ | $-\mathbf{e}_2$ | $\mathbf{e}_1$ | $-\mathbf{e}_7$ | $\mathbf{e}_5$ | $-\mathbf{e}_4$ | **-1** | $\mathbf{e}_3$ |
| $\mathbf{e}_7$ | $\mathbf{e}_7$ | $\mathbf{e}_4$ | $\mathbf{e}_5$ | $\mathbf{e}_6$ | $-\mathbf{e}_1$ | $-\mathbf{e}_2$ | $-\mathbf{e}_3$ | **-1** |

The subspace of octonion space spanned by scaled units that fail to commute with $\mathbf{e}_7$, $\{x_1\mathbf{e}_1, x_2\mathbf{e}_2, x_3\mathbf{e}_3, x_4\mathbf{e}_4, x_5\mathbf{e}_5, x_6\mathbf{e}_6\}$ say, is linearly mapped by an involution given by the matrix:

$$\begin{bmatrix} & 0 & & -i & 0 & 0 \\ & & & 0 & -i & 0 \\ & & & 0 & 0 & -i \\ i & 0 & 0 & & & \\ 0 & i & 0 & & 0 & \\ 0 & 0 & i & & & \end{bmatrix} \qquad (1)$$

This involution gives rise to projections for even and odd parts of bilinear forms in **x** and **y** with the required properties listed in reference [7]. The even part ((a8) in the appendix):

$$\mathbf{x} \bullet \mathbf{y} = x_1y_1+x_2y_2+x_3y_3+x_4y_4+x_5y_5+x_6$$

particularizes (for **y** = **x**) to a Riemannian metric on six-space and the odd part gives:

(i) a vector product
$$\begin{aligned}\mathbf{x} \times \mathbf{y} &= (+x_6y_2-x_5y_3+x_3y_5-x_2y_6)\,\boldsymbol{e}_1 \\ &+ (-x_6y_1+x_4y_3-x_3y_4+x_1y_6)\,\boldsymbol{e}_2 \\ &+ (+x_5y_1-x_4y_2+x_2y_4-x_1y_5)\,\boldsymbol{e}_3 \\ &+ (-x_3y_2+x_2y_3+x_6y_5-x_5y_6)\,\boldsymbol{e}_4 \\ &+ (+x_3y_1-x_1y_3-x_6y_4+x_4y_6)\,\boldsymbol{e}_5 \\ &+ \left(-x_2y_1+x_1y_2+x_5y_4-x_4y_5\right)\boldsymbol{e}_6 \end{aligned} \qquad (a5),$$

and,

(ii) a symplectic form $\mathbf{x} \otimes \mathbf{y} = x_1y_4+x_2y_5+x_3y_6- x_4y_1- x_5y_2 - x_6y_3$ (a9)

## 3 Orientability considerations

A vector $\mathbf{x} = x_1\,\mathbf{e}_1 + x_2\,\mathbf{e}_2 + x_3\,\mathbf{e}_3$ and bivector $\mathbf{y} = y_4\,\mathbf{e}_4, + y_5\,\mathbf{e}_5 + y_6\,\mathbf{e}_6$ may be bi-orthogonal if $\mathbf{x} \times \mathbf{y} = \mathbf{0}$. To particularize this $x_1\,\mathbf{e}_1$ is bi-orthogonal to $y_4\,\mathbf{e}_4$.

The involution $x_1 \rightarrow - x_1$; $y_4 \rightarrow y_4$ yielding $(\mathbf{x} \otimes \mathbf{y})_7 = - (\mathbf{x} \otimes \mathbf{y})_7$ expresses an inner orientation of **x** and an outer orientation of **y**.

At the same time the involution $x_1 \rightarrow x_1$; $y_4 \rightarrow - y_4$ yielding $(\mathbf{x} \otimes \mathbf{y})_7 = - (\mathbf{x} \otimes \mathbf{y})_7$ once again expresses an outer orientation of **x** and an inner orientation of **y**.

R R Birss has pointed out the relevance of the distinction between inner and outer orientations for the handling of improper rotations in condensed matter physics [8,9,10].

It will be noticed that it is the association between an oriented vector and a nonoriented bivector or between an oriented bivector and a nonoriented vector that contributes to chirality as measured by the symplectic product.

## Appendix. Six-dimensional space from eight-dimensional

"The variety of beings should not rashly be diminished" Immanuel Kant.

The conditions for orthogonality and incidence of lines and planes in space, represented as vectors and bivectors, are shown to be expressible in terms of a six-dimensional subalgebra of octonion algebra. The approach depends upon explicit use of the octonion associator and antiassociator.

The present state of understanding in condensed state physics is that certain physical entities having characteristic geometrical attributes are represented by polar or axial vectors. The distinction is according to the transformation properties under improper rotations in space. In other words the two types of vectors have been conflated to the extent that their transformation properties under proper rotations are identical. The present work supports the view that a better distinction than that between polar and axial vectors is one between inner and outer oriented vectors and bivectors. This has been discussed in section 3.Vectors and bivectors contribute separately to observable phenomena in space. In characterizing the algebraic properties of the equations of mathematical physics vectors and bivectors should be handled as representations of objects having distinctive geometrical properties.

It is natural to expect that the three components of a vector and the three components of a bivector should together be taken as the components of a vector in six-dimensional space. It is in this way that Kant's admonishment can be acted on. Plücker, whose interest in the geometry of space [A1] stemmed from consideration of symmetry aspects of condensed state phenomena [A2], and Ball, who addressed degrees of freedom for placement of rigid bodies [A3], followed this approach.

Insofar as a vector spaces may be regarded as representations of algebras those of Plücker and Ball might be thought to present difficulties. Division algebras have sets of units of size one (real numbers), two (complex numbers), four (quaternions) or eight (octonions). However, it transpires that the octonions possess a Lie subalgebra of dimension six by virtue of the existence of a vector (cross) product. It will be shown below that this subalgebra lends itself to the analysis of the geometric properties (both pre-metric and metric) of objects in space.

Consider the octonion:

$$\mathbf{x} = x_0\mathbf{e_0} + x_1\mathbf{e}_1 + x_2\mathbf{e_2} + x_3\mathbf{e_3} + x_4\mathbf{e_4} + x_5\mathbf{e_5} + x_6\mathbf{e_6} + x_7\mathbf{e_7} \qquad \text{(a1)}$$

where the amplitudes $x_i$ are drawn from an appropriate field $\mathbb{F}$ and multiplication is given by:

| | $\mathbf{e_0}$ | $\mathbf{e_1}$ | $\mathbf{e_2}$ | $\mathbf{e_3}$ | $\mathbf{e_4}$ | $\mathbf{e_5}$ | $\mathbf{e_6}$ | $\mathbf{e_7}$ |
|---|---|---|---|---|---|---|---|---|
| $\mathbf{e_0}$ | **1** | $\mathbf{e_1}$ | $\mathbf{e_2}$ | $\mathbf{e_3}$ | $\mathbf{e_4}$ | $\mathbf{e_5}$ | $\mathbf{e_6}$ | $\mathbf{e_7}$ |
| $\mathbf{e_1}$ | $\mathbf{e_1}$ | **-1** | $\mathbf{-e_6}$ | $\mathbf{e_5}$ | $\mathbf{e_7}$ | $\mathbf{-e_3}$ | $\mathbf{e_2}$ | $\mathbf{-e_4}$ |
| $\mathbf{e_2}$ | $\mathbf{e_2}$ | $\mathbf{e_6}$ | **-1** | $\mathbf{-e_4}$ | $\mathbf{e_3}$ | $\mathbf{e_7}$ | $\mathbf{-e_1}$ | $\mathbf{-e_5}$ |
| $\mathbf{e_3}$ | $\mathbf{e_3}$ | $\mathbf{-e_5}$ | $\mathbf{e_4}$ | **-1** | $\mathbf{-e_2}$ | $\mathbf{e_1}$ | $\mathbf{e_7}$ | $\mathbf{-e_6}$ |
| $\mathbf{e_4}$ | $\mathbf{e_4}$ | $\mathbf{-e_7}$ | $\mathbf{-e_3}$ | $\mathbf{e_2}$ | **-1** | $\mathbf{e_6}$ | $\mathbf{-e_5}$ | $\mathbf{e_1}$ |
| $\mathbf{e_5}$ | $\mathbf{e_5}$ | $\mathbf{e_3}$ | $\mathbf{-e_7}$ | $\mathbf{-e_1}$ | $\mathbf{-e_6}$ | **-1** | $\mathbf{e_4}$ | $\mathbf{e_2}$ |
| $\mathbf{e_6}$ | $\mathbf{e_6}$ | $\mathbf{-e_2}$ | $\mathbf{e_1}$ | $\mathbf{-e_7}$ | $\mathbf{e_5}$ | $\mathbf{-e_4}$ | **-1** | $\mathbf{e_3}$ |
| $\mathbf{e_7}$ | $\mathbf{e_7}$ | $\mathbf{e_4}$ | $\mathbf{e_5}$ | $\mathbf{e_6}$ | $\mathbf{-e_1}$ | $\mathbf{-e_2}$ | $\mathbf{-e_3}$ | **-1** |

Table 1 Octonion multiplication

**A1 Vector product on six-space**

The octonion associator:

$$[\mathbf{x}\,, \mathbf{e}_7, \mathbf{y}] = (\mathbf{x}\mathbf{e}_7)\mathbf{y} \; - \mathbf{x}(\mathbf{e}_7\mathbf{y})$$

defines a product $A\colon \mathbb{V} \times \mathbb{V} \rightarrow \mathbb{V}$ on six-space:

$$\mathbf{x} \times \mathbf{y} = \tfrac{1}{2}\pi\mathbf{e_7}[\mathbf{x},\mathbf{e_7},\mathbf{y}] \qquad \text{(a2)}$$

where:

$$\pi\mathbf{x} = x_0\mathbf{e_0} - x_1\mathbf{e}_1 - x_2\mathbf{e_2} - x_3\mathbf{e_3} + x_4\mathbf{e_4} + x_5\mathbf{e_5} + x_6\mathbf{e_6} - x_7\mathbf{e_7} \qquad \text{(a3)}$$

and:

$$\mathbf{e_7x} = -\,x_7\mathbf{e_0} - x_4\mathbf{e}_1 - x_5\mathbf{e_2} - x_6\mathbf{e_3} + x_1\mathbf{e_4} + x_2\mathbf{e_5} + x_3\mathbf{e_6} + x_0\mathbf{e_7} \qquad \text{(a4)}$$

so that:

$$\begin{aligned}\mathbf{x} \times \mathbf{y} = {} & (+x_6y_2-x_5y_3+x_3y_5-x_2y_6)\,\boldsymbol{e}_1 \\ & + (-x_6y_1+x_4y_3-x_3y_4+x_1y_6)\,\boldsymbol{e}_2 \\ & + (+x_5y_1-x_4y_2+x_2y_4-x_1y_5)\,\boldsymbol{e}_3 \\ & + (-x_3y_2+x_2y_3+x_6y_5-x_5y_6)\,\boldsymbol{e}_4 \\ & + (+x_3y_1-x_1y_3-x_6y_4+x_4y_6)\,\boldsymbol{e}_5 \\ & + (-x_2y_1+x_1y_2+x_5y_4-x_4y_5)\,\boldsymbol{e}_6 \end{aligned} \tag{a5}$$

Notice that the set $\{\mathbf{e_1}, \mathbf{e_2}, \mathbf{e_3}, \mathbf{e_4}, \mathbf{e_5}, \mathbf{e_6}\}$ is closed under the vector product and that this set is a basis for a linear vector space over some field, $\mathbb{F}$.

### A2 Inner product on six-space

In addition to the product $\mathbf{x} \circ \mathbf{y}$ defined above the antiassociator gives rise to a complex valued inner product that is made up of a scalar part and a trivector (pseudo-scalar) part. The explicit expression for the octonion antiassociator is:

$$\begin{aligned}\{\mathbf{x}, \mathbf{e_7}, \mathbf{y}\} = {} & 2\,(-x_4y_1-x_5y_2-x_6y_3+x_1y_4+x_2y_5+x_3y_6)\mathbf{e_0} \\ & +2\,(x_1y_1+x_2y_2+x_3y_3+x_4y_4+x_5y_5+x_6y_6)\mathbf{e_7}\end{aligned} \tag{a6}$$

As with the associator, it makes sense to transpose this:

$$\begin{aligned}\mathbf{e_7}\,\pi\{\mathbf{x}, \mathbf{e_7}, \mathbf{y}\} = {} & 2\,(x_1y_1+x_2y_2+x_3y_3+x_4y_4+x_5y_5+x_6y_6)\mathbf{e}_0 \\ & +2\,(x_1y_4+x_2y_5+x_3y_6-x_4y_1-x_5y_2-x_6y_3)\mathbf{e_7}\end{aligned} \tag{a7}$$

but for a different reason. The transpose defines an inner product:

$$AA: \mathbb{V} \times \mathbb{V} \rightarrow \mathbb{F}$$

on what is now Hermitian six-space. The symmetric real part of the inner product of **x** and **y**:

$$\mathbf{x} \bullet \mathbf{y} = x_1y_1+x_2y_2+x_3y_3+x_4y_4+x_5y_5+x_6y_6 \tag{a8}$$

particularizes to a Riemannian metric and the antisymmetric imaginary part:

$$\mathbf{x} \otimes \mathbf{y} = x_1y_4+x_2y_5+x_3y_6-x_4y_1-x_5y_2-x_6y_3 \tag{a9}$$

is a symplectic form. The implication of these emergent properties is that six-space is a linear vector space over either the complex field or the field of Gaussian rationals since it is from one of these fields that the inner product is drawn.

In section **A3** it will be shown that, together, the three products $\times$, $\otimes$ and $\bullet$ allow orthogonality and incidence to be defined in respect of vectors and bivectors. This may facilitate an alternative derivation of Smith's [A4] classification of anisotropic media in terms of the geometry of their wave surfaces.

### A3 Orthogonality and incidence

A preliminary statement of the connection between algebraic and geometric properties is the following:

The vanishing of various products in the algebra corresponds to either:

1) bi-orthogonality of a line and a plane
2) incidence (either in the sense that two lines may be collinear, two planes may be coplanar or that a plane contains a line)
3) orthogonality of two lines or of two planes

Thus, taken together, the products defined in sections **A1** and **A2** allow a resolution of the issue of possible geometric relationships set out above.

It will be noticed that 1) and 2) above are formulations that are pre-metric but do not preclude the introduction of a metric in appropriate circumstances.

The following correspondencies exist between geometric relationships of multivectors and Lie (Kähler) algebraic properties:

1) The vector product, $\times$, (a5) of a vector and a bivector is a vector unless the factors are bi-orthogonal in which case the product is zero.
2) The vector product, $\times$, (a5) of two vectors or of two bivectors is a bivector unless the factors are incident in which case the product is zero.
3) The symplectic product, $\otimes$, (a9) of a mutually bi-orthogonal vector-bivector pair is a trivector.
4) The Riemann product, $\bullet$, (a8) of two vectors or of two bivectors is a scalar unless the two factors are orthogonal in which case the product is zero.

### References (Appendix)